\documentclass[aps,twocolumn,nolongbibliography,preprintnumbers,prl,amsmath,amssymb,amsfonts,superscriptaddress,floatfix,10pt]{revtex4-2}

\usepackage[paperwidth=210mm,paperheight=297mm,centering,hmargin=2cm,tmargin=1.6cm,bmargin=3.cm]{geometry}
\usepackage{booktabs}
\usepackage{graphicx}
\usepackage{amsfonts}
\usepackage{tikz} 
\usepackage{amssymb}
\usepackage{amsmath}
\usepackage{braket}
\usepackage{xcolor}
\usepackage{cancel}
\usepackage{cleveref}

\newcommand{\be}{\begin{equation}}
\newcommand{\ee}{\end{equation}}
\newcommand{\bea}{\begin{eqnarray}}
\newcommand{\eea}{\end{eqnarray}}

\crefname{equation}{Eq.}{Eqs.}
\crefname{figure}{Fig.}{Figs.}
\crefname{table}{Tab.}{Tabs.}

\begin{document}

\title{Crossover and Critical Behavior in the Layered XY Model}

\author{Roman Kracht}
\email{rkracht@phys.ethz.ch}
\affiliation{Institute for Theoretical Physics, ETH Zurich, CH-8093 Zurich, Switzerland}
\author{Andrea Trombettoni}
\affiliation{SISSA and INFN Sezione di Trieste, Via Bonomea 265, I-34136 Trieste, Italy}
\author{Ilaria Maccari}
\affiliation{Institute for Theoretical Physics, ETH Zurich, CH-8093 Zurich, Switzerland}
\author{Nicolò Defenu}
\affiliation{Institute for Theoretical Physics, ETH Zurich, CH-8093 Zurich, Switzerland}
\affiliation{CNR-INO, Area Science Park, Basovizza, I-34149 Trieste, Italy}
\date{\today}

\begin{abstract}
Motivated by the interplay between 2D and 3D scaling signatures observed in unconventional layered superconductors, we present a systematic Monte Carlo study of the three-dimensional classical XY model with anisotropic in-plane $J_{\parallel}$ and inter-plane $J_{\perp}$ couplings. Our study includes very small values of the system anisotropy $\Delta=J_{\perp}/J_{\parallel}$ 
not studied before, and focuses on characterizing the crossover from quasi-2D topological scaling to genuine 3D critical behavior. The numerical results for the critical temperature unambiguously reveal a logarithmic scaling with $\Delta$, directly related to the topological scaling in the 2D limit. Despite the 3D nature of the layered XY criticality, topological scaling signatures survive up to system sizes comparable to the crossover length $\ell_{J}$, which diverges at small $\Delta$ with a scaling behavior reminiscent of the Berezinskii–Kosterlitz–Thouless (BKT) transition. This shows that genuine 3D symmetry-breaking behavior emerges only at exceedingly large system sizes when the anisotropy is very strong. Our results indicate that new experimental evidence is required to clarify the extent to which the critical signatures observed in layered strongly correlated materials are shaped by their pronounced anisotropy.
\end{abstract}

\maketitle

\emph{Introduction:} The initial observation of unconventional superconductivity characterized by high critical temperatures~\cite{bednorz1986possible},
sparked extensive investigations of materials with similar lattice and electronic properties, with the aim of discovering new high-$T_c$ superconducting systems and elucidating their pairing mechanism~\cite{zhou2021high}.
A common feature among these compounds is their layered architecture, where two-dimensional~(2D) sheets hosting strongly correlated electrons are weakly coupled through interstitial oxide structures. 
Although theoretical studies predominantly focus on the physics of these 2D planes~\cite{scalapino2012common,lee2006doping},
high-$T_{c}$ materials exhibit clear evidence of critical scaling behavior consistent with the three-dimensional~(3D) XY universality class~\cite{salamon1993critical,overend1994scaling,kamal1994penetration,pasler1998critical}, with the possible appearance of two 
temperatures for the 2D and 3D behavior~\cite{tinkham2004introduction}.
Interestingly, the 3D critical behavior often coexists with signatures of the Berezinskii–Kosterlitz–Thouless~(BKT) mechanism~\cite{berezinskii1972destruction, kosterlitz1972long, kosterlitz1973ordering,jose2013years}, which is characteristic of 2D systems~\cite{stamp1988kosterlitz,yeh1989quasi,bulaevskii1992fluctuations,pradhan1993observation}. Moreover, the experimental identification of the 2D~versus~3D scaling behavior appears to depend strongly on both the material and the specific sample under investigation~\cite{triscone1990span,gao19933d, wan1996transport, vasyutin2006nonlinearity,li2007two,tranquada2008evidence,baity2016effective,nizhankovskiy2019bkt,guo2020crossover}.

Beyond cuprates, layered architectures have been the subject of long-lasting interest. A layered superconductor was realized back in the 60's by alternating layers of graphite and alkali metals~\cite{hannay1965superconductivity}, followed by the discovery
of naturally occurring compounds of transition-metal
dichalcogenide layers intercalated with
organic, insulating molecules~\cite{gamble1970superconductivity} and by the creation of artificial samples with alternating layers of different
metals with different transition temperatures~\cite{ruggiero1980superconductivity}.
Layered superconductors have subsequently emerged across a broad and rapidly expanding family of materials~\cite{mizuguchi2016recent, mizuguchi2018layered}, including iron-based superconductors~\cite{kamihara2008iron, takahashi2008superconductivity, paglione2010high, stewart2011superconductivity, meng2024layerdependent}, infinite-layer nickelates~\cite{li2019superconductivity, norman2020entering, pickett2021the, sun2023signatures, wang2025recent}, and finally
superconducting van~der~Waals heterostructures~\cite{geim2013der,
novoselov2016materials,castellanosgomez2022der} which offer unprecedented control over the effective dimensionality and interlayer coupling, making the 2D~to~3D crossover directly tunable.

Apart from superconducting systems,
this structural motif plays a central role across a wide range of physical platforms, spanning magnetic materials~\cite{huang2017layerdependent, blei2021synthesis, jiang2021recent, wang2022magnetic,huang2023pressurecontrolled, ilyas2025layered, algarni2025recent}, molecular crystals~\cite{watson2001magnetic,zvyagin2007magnetic,jeong2017magnetic}, 
as well as ultracold atoms in 1D optical lattices, where the tunneling between quasi-2D or pancake-like systems can be precisely controlled~\cite{cataliotti2001josephson, ho2004deconfinement, sommer2012evolution, iazzi2012anisotropic, sajna2014dynamical}.

Theoretical understanding and guidance of these experimental platforms require a solid understanding of universal scaling and critical behavior in layered structures with~$U(1)$ symmetry. Although the physics of the dimensional crossover is rather natural in most models~\cite{binder1992wetting}, the XY model, the paradigmatic $U(1)$ effective theory for superconducting phase fluctuations, is special: the nature of its phase transition changes drastically between two and three dimensions. In~2D, it undergoes the topological BKT transition, beyond the standard Ginzburg-Landau framework.
Constrained by the Mermin-Wagner theorem~\cite{mermin1966absence}, the system has zero magnetization and yet it develops a quasi-long-range order at low temperatures, destroyed at the critical point by the unbinding of vortex-antivortex pairs.
In contrast, the 3D~XY~model undergoes a conventional second-order transition, where the low-temperature magnetized phase melts due to the proliferation of vortex loops above $T_{c}$~\cite{williams1987vortex,shenoy1995anisotropic}.

To address this dimensional crossover, early theoretical studies 
were developed on the anisotropic 3D~XY~model~\cite{berezinskii1973thermodynamics,pokrovskii1973magnetic,hikami1980phase,chui1988monte, janke1990crossover,minnhagen1991monte,schmidt1992dimensional,pierson1992critical,pierson1992renormalizationgroup,weber1992monte,baeriswyl1992ordering,fischer1993kosterlitzthouless,pierson1995critical,friesen1995vortex}, the Lawrence-Doniach model in which 2D~$\psi^4$ models are coupled by a Josephson coupling term~\cite{lawrence1970theory,tinkham2004introduction} and the Lawrence-Doniach coupled sine-Gordon models~\cite{pierson1992critical,pierson1995it}, whose phenomenology at low energy is closely related~\cite{pierson1994mapping,nandori2007applicability}.

Despite experimental observations of two critical temperatures in high-$T_{c}$ superconductors~\cite{tinkham2004introduction}, one for the onset of in-plane superconductivity and the other for bulk superconductivity~\cite{nizhankovskiy2019bkt}, most theoretical pictures suggest that the layered XY model possesses a \textit{single} critical line as a function of the ratio between inter-plane and in-plane ferromagnetic couplings ($\Delta = J_\perp / J_\parallel$). At the critical line $T_{c}(\Delta)$, the model undergoes a second-order phase transition, which lies in the 3D XY universality.
The presence of a single critical line in the layered 3D XY model may still describe experimental observations due to the presence of an extended crossover region above the line $\varepsilon \sim \Delta^{-1/\nu}$, with $\varepsilon = |T - T_{c}(\Delta)| / T_{c}(\Delta)$ and $\nu$ being the 3D~XY (thermal) correlation length critical exponent. In this region,
the system effectively exhibits BKT-like scaling. Equivalently, there exists a crossover length $\ell_{J}$ below which the model appears two-dimensional and displays BKT critical behaviour, while genuine three-dimensional criticality emerges only at much larger scales~\cite{rancon2017kosterlitzthouless}.

 This theoretical scenario differs fundamentally from Friedel’s early proposal~\cite{friedel1988quasi}, in which 2D layers decouple through the proliferation of fluxons between planes, a mechanism that arises in models where the inter-plane coupling extends beyond the simple Josephson form~\cite{dzierzawa1996friedel}. The intricate interplay between these theoretical and experimental pictures still lacks a correspondingly well‑developed and unified numerical understanding. Numerical evidence and study of the 3D~layered~XY model has remained surprisingly limited and has been restricted mainly to small system sizes~\cite{chui1988monte, tanner1994dimensionalitats}. Moreover, the characteristic \textit{Josephson} length scale governing the dimensional crossover from 2D~BKT-like physics to true 3D~behaviour has never been explicitly studied. This lack of numerical evidence has prevented the layered 3D~XY~model from becoming broadly recognized as the effective description of the critical behaviour of layered unconventional  superconductors. In the following, we are going to bridge this gap and argue that the model could still serve as proper description of critical scaling in layered materials.

\emph{The Model:} The 3D~anisotropic~XY~model describes a system of weakly coupled 2D~XY~layers, formed by ferromagnetically interacting planar spins. The Hamiltonian of the model is given by
\be
\mathcal{H} = - J_{\parallel} \sum_{\langle ij \rangle_\parallel} \vec{s}_i \cdot \vec{s}_j - J_{\perp} \sum_{\langle ij \rangle_\perp} \vec{s}_i \cdot \vec{s}_j \label{eq:xy_hamiltonian},
\ee
where $\vec{s}_i \equiv (\cos \theta_i, \sin \theta_i)$ is the planar spin on the site $i$, $\langle ij \rangle_\parallel$ indicates the sum over in-plane nearest neighboring and $\langle ij \rangle_\perp$ over inter-plane nearest neighboring spins. In the thermodynamic limit $L\to\infty$, both the in-plane dimensions and the number of coupled layers diverge, yielding a truly 3D system rather than a quasi-2D one. This setting differs crucially from recent Monte Carlo studies that either keep the number of layers fixed~\cite{masini2025helicity} or couple adjacent layers only through a single perpendicular plane~\cite{hu2025longrange}.

Two limiting cases anchor the phase diagram of the model. For~${\Delta=1}$, the model reduces to the isotropic 3D~XY~model, which exhibits a conventional second-order phase transition, associated with the spontaneous breaking of the $U(1)$ global symmetry. The critical temperature and exponents in this case are known up to high accuracy~\cite{xu2019highprecision}.
On the other hand, for ${\Delta=0}$ the planes decouple and one recovers $L$ independent 2D~XY~layers of dimension $L\times L$. Each layer undergoes the well-known BKT transition at $T_{\mathrm{BKT}}/J_{\parallel} \approx 0.8929(4)$~\cite{hasenbusch2005twodimensional}.

Between these two limiting cases, previous studies~\cite{hikami1980phase,chui1988monte} proposed that the critical temperature of the 3D~anisotropic~XY~model scales logarithmically as a function of~$\Delta$:
\be
    \left[ T_c (\Delta) - T_{BKT} \right]^{-1/2} = \alpha - \beta \ln \Delta \label{eq:crit_temp_delta},
\ee
where $\alpha$ and $\beta$ are non-universal coefficients. This functional form has been obtained by considering a Lawrence-Doniach model, where each XY~plane is represented as an effective sine-Gordon model~\cite{hikami1980phase,benfatto2007kosterlitz} and has been verified in early Monte Carlo studies~\cite{chui1988monte}, which have partially remained unpublished~\cite{tanner1994dimensionalitats}.

In order to validate the current theoretical picture of the layered~XY~model, one must first verify whether Eq.~\eqref{eq:crit_temp_delta}, which was actually obtained for coupled sine-Gordon models, 
holds over the entire $\Delta$~range. Indeed, the logarithmic scaling of the critical temperature is characteristic of coupled systems governed by topological scaling. By contrast, the traditional mean-field calculation on Hamiltonian~\eqref{eq:xy_hamiltonian} shows that the order parameter $m$ satisfies
$m = \frac{I_1\!\left[(4J_\parallel + 2J_\perp)\, m/k_B T\right]}{I_0\!\left[\beta (4J_\parallel + 2J_\perp)\, m/k_B T\right]}$, where~$I_n(x)$ denotes the $n$-th modified Bessel function~\cite{abramowitz1964handbook}. This leads to
$\frac{T_c(\Delta)}{T_c(\Delta=0)} = 1 + \frac{\Delta}{2}$,
which is consistent with the results for layered models with spontaneous symmetry breaking~\cite{stanley1971introduction}, where $T_c(\Delta) \propto \Delta^{1/\gamma}$ and the mean-field exponent satisfies $\gamma = 1$. However, as expected, this conventional symmetry-breaking picture disagrees with the topological-scaling prediction of Eq.~\eqref{eq:crit_temp_delta}.

In this Letter, we report the first large-scale Monte Carlo investigation of the layered 3D XY model and provide quantitative validation of the three key predictions of the crossover scenario:
\begin{enumerate}
\item The scaling of the transition temperature is accurately described by Eq.~\eqref{eq:crit_temp_delta}, even for the full XY~model.
\item The transition exhibits genuine 3D critical behavior throughout the entire $\Delta$~range.
\item BKT scaling persists for system sizes $\lesssim \ell_{J}$.
\end{enumerate}

\emph{Results:}
We simulate the 3D~layered~XY~model~\eqref{eq:xy_hamiltonian} using heat bath updates in the thermalization process~\cite{miyatake1986the} aided by parallel-tempering swaps~\cite{swendsen1986replica}, while during measurements we interleave the heat bath with Wolff-cluster steps~\cite{swendsen1987nonuniversal,wolff1989collective} to reduce the autocorrelation time.
For all simulations, we use the blocked bootstrap resampling method to propagate statistical errors. More practical details are provided in the Supp.~Mat.~\cite{supplemental}. In order to locate the critical point, we compute the bulk magnetization, ${\vec{M} = L^{-3}\sum_i \vec{s}_i = M\,e^{i\phi}}$ and the corresponding Binder cumulant:
\be
b = \langle M^4 \rangle / \langle M^2 \rangle^2 ,
\label{binder}
\ee
which, for second-order phase transitions, follows the finite-size scaling form $f_b$~\cite{binder1981finite}:
\be
b(T,L)=f_b\!\left[(T-T_c)L^{1/\nu}\right] \label{eq:binder_scaling}.
\ee
As a result, $b$~takes a universal value at the critical temperature~$T_c$, up to weak scaling corrections, leading to characteristic crossings of curves for different system sizes~$L$.
We also compute the 3D~superfluid stiffness~$\rho_s$. That is a non-local quantity characterizing the change in free energy~$F$ under a phase twist~$\phi$ of the boundary conditions along a given direction:
\be
    \rho_s = \frac{1}{L^3} \left. \frac{\partial^2 F(\phi)}{\partial^2 \phi} \right|_{\phi=0}.
\ee
In 3D, it has scaling dimension ${[\rho_s]=L^{-1}}$~\cite{gottlob1993critical}; therefore, the rescaled quantity ${j = L\rho_s}$ follows a similar scaling law as Eq.~\eqref{eq:binder_scaling} with scaling function $f_j\!\left[(T-T_c)L^{1/\nu}\right]$ and provides an independent way of locating the critical point of the model.
\begin{figure}[t]
    \centering
    \includegraphics[width=\linewidth]{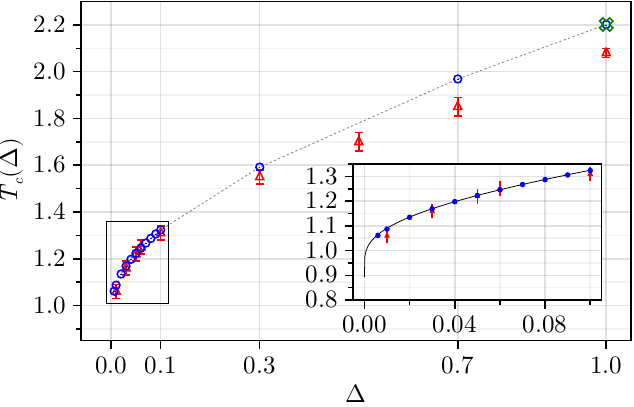}
    \caption{Critical temperature~$T_c$ as a function of the coupling ratio~${\Delta = J_\perp / J_\parallel}$: The blue dots show our Monte Carlo results (errors smaller than markers) with the gray line being a guide to the eye. The red triangles show results from~\cite{tanner1994dimensionalitats} and the green cross shows the isotropic high-precision result from~\cite{xu2019highprecision}. The solid black line indicates the fit with Eq.~\eqref{eq:crit_temp_delta}, where $T_{BKT}$ is kept fixed to the literature value~$0.89298(4)$~\cite{hasenbusch2005twodimensional}.}
    \label{fig:critical_temperature}
\end{figure}

In practice, $T_c(\Delta)$ is found by expanding the respective scaling functions of $b$ and $j$ as a Taylor series in $(T-T_c) L^{1/\nu}$ and fitting them to the Monte Carlo data. Rather than performing a single fit across all sizes, we group the data points in sets~$(L, 2L)$ and fit~$T_c$ separately for each, thus giving us an effective size-dependent critical temperature~$T_c^{\mathrm{eff}} (L)$.
The thermodynamic limit result is  estimated for each anisotropy~$\Delta$, with a power-law ansatz ${T_c^{\mathrm{eff}} (L) = T_c + \mu L^{-p}}$, treating $T_c$, $\mu$ and $p$ as free parameters. All values of $T_c^\mathrm{eff} (L)$ and the associated extrapolation fits are presented in the Supp.~Mat.~\cite{supplemental}. The final result for~$T_c (\Delta)$ is shown in Fig.~\ref{fig:critical_temperature}, where we have combined independent extrapolations obtained from the Binder cumulant and rescaled stiffness.

By fitting our numerical results for $T_c (\Delta)$ with the logarithmic scaling in Eq.~\eqref{eq:crit_temp_delta} in the range ${0 < \Delta \leq 0.08}$, with $T_{BKT}$ fixed to the literature value $0.8929(4)$~\cite{hasenbusch2005twodimensional}, we obtain $\chi^2 /\mathrm{DOF} \approx 1.28$ with fitting parameters $\alpha = 0.792(2)$ and $\beta = 0.317(1)$ in good agreement with the Lawrence-Doniach coupled sine-Gordon models~\cite{pierson1992critical,pierson1995it} and the real space RG picture~\cite{chattopadhyay1994kosterlitzthouless,shenoy1995anisotropic}. Furthermore, keeping $T_{BKT}$ as a free fit parameter yields a value of~$0.874(7)$ that deviates from the value of the purely 2D~literature by less than $3\%$. In order to improve this precision, the numerical simulations need to reach exponentially small values of~$\Delta$, where, however, the size needed to reach the scaling regime also diverges, as argued in the following.
\begin{figure}
    \includegraphics[width=\linewidth]{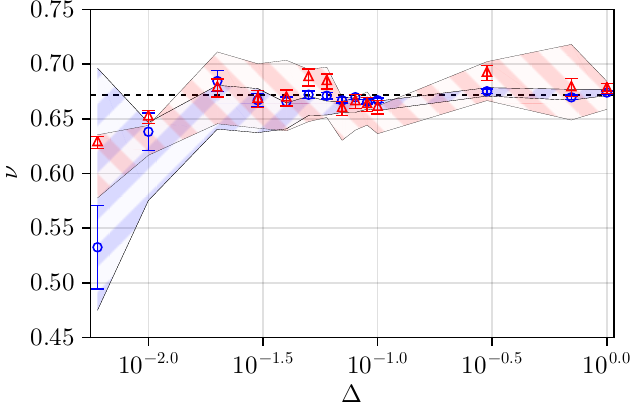}
    \caption{Correlation length critical exponent~$\nu$ as a function of the coupling ratio~$\Delta = J_\perp / J_\parallel$:
    The blue dots represent extrapolation results from the Binder cumulant~$b$ and the red triangles those from the rescaled stiffness~$j$. In contrast, the average of the $5$ largest system sizes (without extrapolation) is indicated by the blue $\diagup$-shaded area for~$b$ and by the red $\diagdown$-shaded area for~$j$. The horizontal dashed line represents the isotropic high-precision result from~\cite{xu2019highprecision}.
    }
    \label{fig:nu_over_delta}
\end{figure}

In summary, the numerical estimates of~$T_{c}(\Delta)$ reported in Fig.~\ref{fig:critical_temperature} show excellent agreement with the prediction in Eq.~\eqref{eq:crit_temp_delta} as well as with points (1.)~and~(2.) of our
three predictions. To further scrutinize hypothesis~(2.), we repeat the full analysis under the assumption of BKT~scaling. The relative quality of the 3D~XY versus BKT~scaling hypotheses is assessed through a $\chi^{2}$-test, whose results are presented in the End Matter (see Fig.~\ref{fig:chi2red_comparison_binder}). As anticipated, the data clearly violates the BKT scaling form for~${\Delta \lesssim 1}$, where the behavior is unmistakably consistent with 3D~XY~criticality. This 3D-XY-dominated region extends to approximately $\Delta \simeq 0.04$, below which the BKT~hypothesis becomes progressively less incompatible with the numerical results. Nevertheless, across the entire explored $\Delta$~range, the 3D~XY universality class remains systematically favored, as highlighted in the inset of Fig.~\ref{fig:chi2red_comparison_binder}.

Consistent with the $\chi^{2}$-analysis, the estimates of the correlation length critical exponent~$\nu$, extracted from the scaling of the Binder cumulant (see End Matter for more details), reproduce the known value of the 3D~XY~universality class in the range~${\Delta\gtrsim 0.01}$ (see the blue points in Fig.~\ref{fig:nu_over_delta}). The results obtained from the scaling of the superfluid stiffness display larger fluctuations even at larger~$\Delta$ (see the red points in Fig.~\ref{fig:nu_over_delta}). At very small~${\Delta \lesssim 0.01}$, our estimates of~$\nu$ start to deviate from the 3D~XY~value. We attribute this behavior to amplified finite-size fluctuations in this regime. This interpretation is supported by the width of the shaded bands in Fig.~\ref{fig:nu_over_delta}, which reflect the increasing spread of the finite-size estimates~$\nu_{\mathrm{eff}} (L)$.
\begin{figure*}
    \centering
    \includegraphics[width=\textwidth]{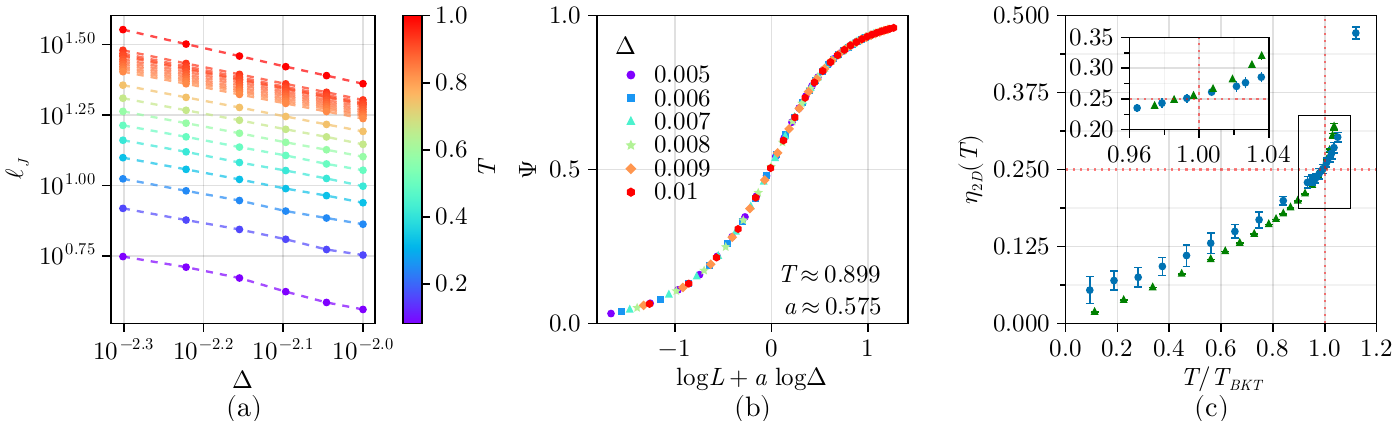}
    \caption{
    \textbf{(a)}~Josephson length scale~$\ell_J$ as a function of the coupling ratio~${\Delta = J_\perp / J_\parallel}$ for various temperatures~$T$. We computed~$\ell_J$ from Eq.~\eqref{eq:josephson_criterion} with a threshold~$t=0.7$. For each temperature, the data points follow straight lines in $\log$-$\log$-scale, thus confirming the power-law scaling from Eq.~\eqref{eq:josephson_length_scaling}.
    \textbf{(b)}~Curve collapse of layer-alignment~$\Psi$ for temperature~${T \approx 0.899}$. We plot~$\Psi$ over the rescaled system size~${\log [ L \Delta^{a (T)} ]}$ for~${L\geq 4}$, where the exponent~${a\approx0.575}$ was determined by minimization of the collapse error.
    \textbf{(c)}~Anomalous dimension~$\eta_{2D}$ over temperature~$T$. The circles indicate curve collapse results from the layer-alignment~$\Psi$ of the 3D~model within a $1\sigma$ confidence interval. The triangles indicate high-precision results from~\cite{maccari2017broadening} from 2D~simulations. The red dotted lines indicate the exact theory prediction $\eta_{2D}(T_{BKT})= 1/4$. All temperatures are normalized by~${T_{BKT} \approx 0.8929(4)}$~\cite{hasenbusch2005twodimensional}.
    }
    \label{fig:josephson}
\end{figure*}

The different nature of the phase transitions at ${\Delta=0}$ (infinite order transition) and ${\Delta=1}$ (spontaneous symmetry breaking) corresponds to qualitatively different mechanisms, i.e. vortex unbinding versus vortex-loop proliferation. Hence, the model displays a pronounced dimensional crossover in the ordered phase. The crossover inherits several characteristics of the BKT phase, in particular the \textit{Josephson length scale} $\ell_J$~\cite{rancon2017kosterlitzthouless}, which controls the flow from effective 2D~BKT-like behavior at intermediate scales $L < \ell_J$ to asymptotic 3D~XY criticality for $L > \ell_J$. The Josephson length is expected to follow a power-law scaling~\cite{rancon2017kosterlitzthouless}:
\be
\ell_J(\Delta,T)\sim \Delta^{-a(T)},\quad a(T)=\big[\,2-\eta_{2D}(T)\,\big]^{-1} \label{eq:josephson_length_scaling},
\ee
where~$\eta_{2D}(T)$ is the anomalous dimension of a strictly 2D~XY~model in the BKT phase. This is defined in terms of spin correlations:
${\langle \cos(\varphi_{\vec 0}-\varphi_{\vec r})\rangle \sim r^{-\eta_{2D}(T)}}$, as~${r\to\infty}$ and establishes a remarkable quantitative connection between 2D and 3D~XY~physics. In order to validate this expectation, it is convenient to consider the magnetization of a single layer~$l$, defined as ${\vec{m}_l = L^{-2} \sum_x \sum_y \vec{s}_{(x,y,l)} = m_l \cdot e^{i \phi_l}}$.
As the system crosses over from quasi-2D to 3D~behavior with increasing system size, the layers in the ordered phase (${T<T_c}$) first develop independent magnetization (${m_l>0}$), which gradually establishes long-range coherence across the stack, resulting in a finite bulk magnetization ${M>0}$ in the thermodynamic limit. This was observed in previous small-scale simulations~\cite{chui1988monte}.
We quantify inter-layer phase locking in terms of the layer-alignment parameter~$\Psi$, which measures the average relative orientation in the magnetization of adjacent layers:
\be
\Psi = \left\langle \frac{1}{L} \sum_l \frac{\vec{m}_{l+1}}{m_{l+1}} \cdot \frac{\vec{m}_l}{m_l} \right\rangle = \left\langle \frac{1}{L} \sum_l \cos (\phi_{l+1} - \phi_l ) \right\rangle . \label{eq:layer_alignment}
\ee

By construction ${0 \leq \Psi \leq 1}$: values ${\Psi \approx 0}$ indicate mutual misalignment (quasi-2D behavior) whereas ${\Psi \approx 1}$ reflects phase locking among layers and thus genuine 3D~order. In the 2D~to~3D crossover, $\Psi$~interpolates smoothly between the limits (see the Supp.~Mat.~\cite{supplemental}) and we can define the Josephson length scale $\ell_J$ by imposing a threshold $t$:
\be
\Psi\!\left( L = \ell_J, \Delta , T \right) \overset{!}{=} t, \qquad 0 < t < 1. \label{eq:josephson_criterion}
\ee
Fig.~\ref{fig:josephson}(a) shows the resulting~$\ell_J$ obtained for threshold ${t=0.7}$ at different coupling ratios~$\Delta$ and temperatures~$T$.
In a $\log$-$\log$~plot, the data points for each~$T$ lie on straight lines, confirming the power-law scaling predicted by Eq.~\eqref{eq:josephson_length_scaling}. Small deviations appear at low temperatures (e.g.~${T \sim 0.2}$), where the crossover occurs at such short~$\ell_J$ that even smaller inter-layer couplings would be required to fully expose the asymptotic scaling. Closer to~$T_{BKT}$, the power law is essentially perfect.

Taken together, Eqs.~\eqref{eq:josephson_length_scaling} and~\eqref{eq:josephson_criterion} imply a single-parameter scaling~$f_{\Psi}$ for the layer-alignment~$\Psi$:
\be
    \Psi\!(L, \Delta, T) = f_{\Psi}\!\left[ L \cdot \Delta^{a(T)} \right], \label{eq:layer_alignment_scaling}
\ee
since inverting~$f_\Psi$ for any fixed threshold~${0 < t <1}$ yields the expected scaling~$\ell_J = f_\Psi^{-1} (t) \cdot \Delta^{-a(T)}$. Eq.~\eqref{eq:layer_alignment_scaling} shows that the choice of~$t$ merely fixes a nonuniversal prefactor of~$\ell_J$ and does not affect the scaling exponent~$a(T)$. As a consequence, all~$\Psi$ curves corresponding to a fixed exponent~$a(T)$ collapse onto a single scaling curve, as shown in Fig.~\ref{fig:josephson}(b) for the case~$T = 0.8993$. This successful collapse provides direct evidence for the validity of Eq.~\eqref{eq:layer_alignment_scaling}. The optimal collapse exponent, ${a \approx 0.575}$ for ${T = 0.8993}$, is then obtained by minimizing an error function adapted from~\cite{bhattacharjee2001measure}. A detailed account of the procedure, together with additional results at other temperatures, is presented in the Supp.~Mat.~\cite{supplemental}.

By systematically minimizing the collapse error, the scaling exponent~$a(T)$ is obtained over a wide range of temperatures.
Then, we obtain estimates of the 2D anomalous dimension~$\eta_{2D} (T)$ from the collapse exponent~$a(T)$. Fig.~\ref{fig:josephson}(c) shows our estimates together with previous high-precision results from~\cite{maccari2017broadening} that were obtained for a single XY~plane with linear size~${L=256}$.
Extracting the 2D~anomalous dimension from the Josephson-length scaling  in the anisotropic 3D~model provides accurate estimates of~$\eta_{2D}$ of the genuine 2D~XY~model over the entire range ${0 < T < T_{BKT}}$. In particular, at ${T=0.886522}$ we obtain ${\eta_{2D} = 0.251(7)}$, which is consistent with the exact result $\eta(T_{BKT})=1/4$. Since we infer~$\eta_{2D}$ from the dimensional crossover of the 3D~model, our uncertainties are naturally larger than those of the dedicated 2D~simulation. The observation of an extended 2D~scaling region, where the spins display in plane coherence, but inter-plane coherence still needs to be established, is consistent with experimental observation of 2D~transport before the onset of 3D~superconductivity in finite samples~\cite{wan1996transport,tranquada2008evidence}.

In summary, the layer-alignment~$\Psi$ provides a clean, dimensionless probe of the ${\mathrm{2D} \to \mathrm{3D}}$~crossover. Its single-parameter scaling captures the Josephson length scale~$\ell_J$, and the resulting curve collapse yields an exponent~$a(T)$ that inherits the BKT scaling of the in-plane correlations~\cite{rancon2017kosterlitzthouless}. The relevance of~$\Psi$ as a practical diagnostic for the crossover is not limited to numerical simulations or to the XY~model. We propose~$\Psi$ as a general tool for analyzing crossover phenomena in layered systems, and we expect that its experimental measurement in real materials may help clarify the role of layering in unconventional superconductors. 

\emph{Conclusions:} Motivated by the complex interplay between 2D~and~3D scaling signatures in real samples and unconventional superconductors, we present a comprehensive Monte Carlo analysis of the 3D~XY~model with anisotropic couplings, targeting (i)~the dimensional crossover from quasi-2D BKT-like behavior to the asymptotic 3D~regime, and (ii)~how the coupling ratio~$\Delta$ controls the extension of the thermodynamic critical region. The critical temperature curve is found to reproduce the expectation from the layered sine-Gordon (see Eq.~\eqref{eq:crit_temp_delta}) and an analysis of the scaling behavior confirms 3D~spontaneous symmetry breaking over the entire $\Delta$~range we analyzed (see Fig.~\ref{fig:nu_over_delta}).

To track the crossover, we introduce the layer-alignment~$\Psi$, which is straightforward to apply in both theoretical analyses and experimental studies of layered systems. Using~$\Psi$, we extract the characteristic crossover (Josephson) length~$\ell_J$ and confirm the power-law behavior predicted by RG~arguments, Eq.~\eqref{eq:josephson_length_scaling}~\cite{rancon2017kosterlitzthouless}. Our results suggest that the coexistence of 2D~and~3D features observed in the thermodynamic and transport properties of real materials hosting unconventional superconducting phases may still originate from their layered structure. Although this structure does not induce a genuinely new phase, it generates a sharp crossover in which 2D~signatures persist with surprising accuracy. To assess the validity of the crossover scenario against alternative proposals involving genuinely new thermodynamic phases~\cite{hern1999sliding}, it is essential to perform additional experimental measurements on different samples and across a broader range of observables. Our results provide a coherent framework for interpreting such future experiments and for characterizing the crossover behavior in layered~$O(2)$ systems.

\section{Acknowledgements}

Discussions with N. Dupuis, I. Pasqua, S.R. Shenoy and G.A. Williams are gratefully acknowledged. This research was funded by the Swiss National Science Foundation (SNSF) grant numbers 200021--207537, 200021--236722 and via the SNSF postdoctoral Grant No. TMPFP2\_217204. Financial support by the Deutsche Forschungsgemeinschaft (DFG, German Research Foundation) under Germany's Excellence Strategy EXC2181/1-390900948 (the Heidelberg STRUCTURES Excellence Cluster) and the Swiss State Secretariat for Education, Research and Innovation (SERI) is also acknowledged.

\bibliography{ref.bib}

\appendix*
\clearpage
\onecolumngrid

\vspace{1em}
\begin{center}
\textbf{\large End Matter}
\end{center}
\vspace{1em}

\twocolumngrid

\begin{figure}[t!]
    \centering
    \includegraphics[width=\linewidth]{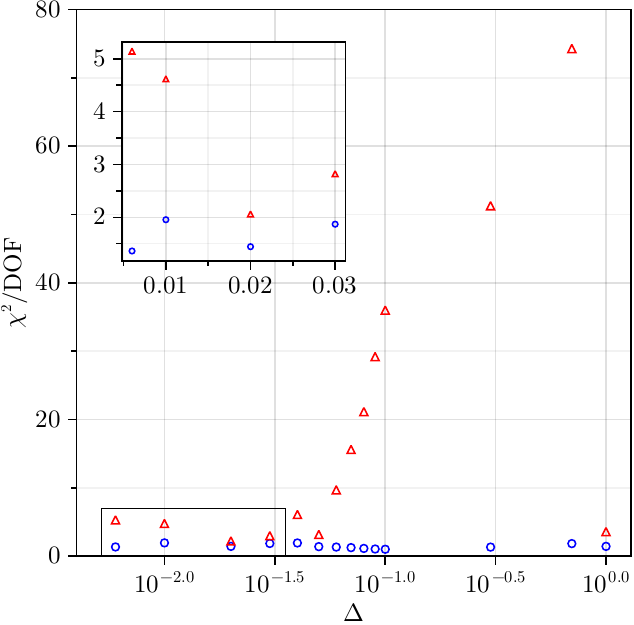}
    \caption{Goodness-of-fit~$\chi^2/\mathrm{DOF}$ for Binder cumulant as a function of the coupling ratio~${\Delta = J_\perp / J_\parallel}$: The blue dots correspond to the second-order finite-size scaling ansatz from Eq.~\eqref{eq:binder_scaling}, whereas the red triangles correspond to the BKT~scaling. Both fits have been performed with the critical temperature and~$\nu$ respectively~$c$ unconstrained. Each scaling function has been expanded as a Taylor polynomial with~$6$ free coefficients.}
    \label{fig:chi2red_comparison_binder}
\end{figure}
\emph{Second-order versus BKT scaling:} Here, we provide numerical evidence that the critical behavior of Hamiltonian~\eqref{eq:xy_hamiltonian} belongs to the 3D~XY universality class. In analogy with previous studies~\cite{gupta1988phase,gupta1992critical,janke1993high} that numerically established the BKT scaling for the XY~model in two dimensions. For second-order phase transitions the correlation length~$\xi$ shows power law divergence with a critical exponent~$\nu$ when approaching the critical temperature: $\xi \sim |T-T_c|^{-\nu}$. For the BKT transition on the other hand, the correlation length diverges as~$\xi \sim e^{c / \sqrt{|T-T_{BKT}|}}$ with a constant~$c$ as~$T_{BKT}$ is approached from above. As a consequence, the Binder cumulant close to a second order phase transition follows Eq.~\eqref{eq:binder_scaling} at leading order, while in the BKT case this scaling law should be replaced by $b(T,L) = f_b (|T-T_c| \cdot (\ln L + c)^2)$. To confirm that the transition is indeed of second order we fit our data for the binder cumulant with both scaling forms, keeping the critical temperature and~$\nu$ or~$c$ respectively as free parameters. 

Fig.~\ref{fig:chi2red_comparison_binder} shows the resulting~$\chi^2/\mathrm{DOF}$ for each value of~$\Delta$, where we have expanded each scaling function as a Taylor polynomial with~$6$ independent coefficients. The second-order fits yield a~$\chi^2 /\mathrm{DOF}$ close to unity. However, the BKT fit consistently yields higher values over the whole range of~$\Delta$, and one can, thus, conclude that second-order scaling is favored. We note that the BKT~$\chi^2 / \mathrm{DOF}$ drops as~${\Delta \to 0}$, another feature that highlights the onset of the dimensional crossover. The full data is tabulated in the Supp.~Mat.~\cite{supplemental}, both for the Binder cumulant and superfluid stiffness. We note that the exact numerical values also implicitly depend on the temperature range chosen for the simulation. Indeed, at the isotropic point~$\Delta=1$, where the critical point is known to high accuracy, both~$\chi^{2}$ values are more compatible than at other values of~$\Delta$. Lastly, we show in the Supp.~Mat.~\cite{supplemental}, that the extracted values of~$\chi^2 /\mathrm{DOF}$ remain stable for increasing order of the Taylor polynomial that approximates the scaling functions.

\begin{figure}[t!]
    \centering
    \includegraphics[width=\linewidth]{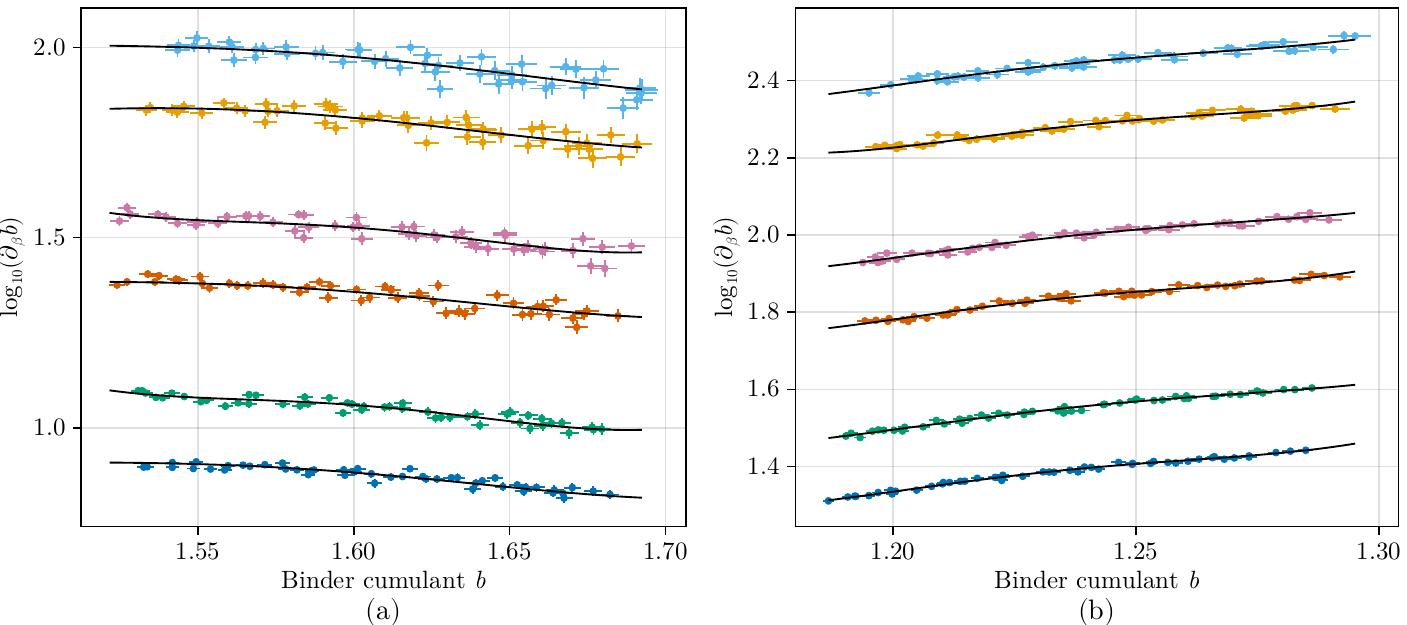}
    \caption{Derivative of Binder cumulant~$b$ with respect to inverse temperature~$\beta = 1/T$ for \textbf{(a)}~coupling ratio~${\Delta = 0.05}$ and \textbf{(b)} isotropic couplings~${\Delta = 1}$. Data points are indicated by circles and the associated fits are indicated by the solid lines. For clarity, here we only show results for system sizes~$L=72,56,36,28,18,14$ (from top to bottom). The full extrapolation used to obtain the results displayed in Fig.~\ref{fig:nu_over_delta} however uses more intermediate system sizes, as explained in the Supp.~Mat.~\cite{supplemental}.}
    \label{fig:binder_log_deriv}
\end{figure}

\emph{Correlation length critical exponent:}
Having established that the phase transition is of second-order for any~$\Delta > 0$, we now describe the method of extracting the correlation-length critical exponent~$\nu$. Let us first note that, in the critical region, the scaling function~$f_b$ in Eq.~\eqref{eq:binder_scaling} is invertible, i.e. we may express the scaling argument in terms of the Binder cumulant: $(T-T_c) \cdot L^{1/\nu} = {f_b}^{-1} (b)$. Close to~$T_c$ we can thus parametrize the temperature-derivative of the Binder cumulant through the Binder cumulant itself:
\begin{equation}
    \partial_T b = L^{1/\nu} f_b^\prime \left( {f_b}^{-1} (b) \right) =: L^{1/\nu} h_b(b).
\end{equation}
This procedure, introduced in~\cite{beiming2023}, allows us to eliminate~$L$ from the argument of the scaling function, with the price that we now have to compute both the Binder cumulant and its temperature-derivative, which however can be done straightforwardly by sampling expectation values~$\langle \mathcal{H} \rangle$, $\langle m^2 \mathcal{H} \rangle$ and $\langle m^4 \mathcal{H} \rangle$ in the Monte Carlo simulation. To fit~$\nu$, we use the logarithm:
\begin{equation}
    \log \partial_\beta b = \frac{1}{\nu} \log L + \log h_b (b)\label{eq:binder_deriv_log}.
\end{equation}
Note that here we use the derivative with respect to inverse temperature~$\partial_\beta = (\partial T / \partial \beta) \partial_T$ for numerical convenience (otherwise one needs take the logarithm of the modulus~$|\partial_T b|$).
Since we perform simulations very close to the critical temperature, the term $\log h_b (b)$ is nearly constant and expanding it as a 4th~order Taylor polynomial is sufficient. We fit Eq.~\eqref{eq:binder_deriv_log} to the Monte Carlo measurements, grouping data in pairs of~$(L,2L)$, as shown in Fig.~\ref{fig:binder_log_deriv} for~${\Delta = 0.05}$ and for the isotropic point~${\Delta = 1}$. Lastly, we extrapolate the effective $L$-dependent critical exponent to the thermodynamic limit for each~$\Delta$ individually. The result is shown in Fig.~\ref{fig:nu_over_delta}.
\onecolumngrid

\clearpage
\onecolumngrid
\setcounter{equation}{0}
\setcounter{figure}{0}
\setcounter{table}{0}
\setcounter{section}{0}
\renewcommand{\theequation}{S\arabic{equation}}
\renewcommand{\thefigure}{S\arabic{figure}}
\renewcommand{\thetable}{S\arabic{table}}
\renewcommand{\thesection}{S\arabic{section}}

\vspace{1em}
\begin{center}
\textbf{\large Supplemental Material}
\end{center}
\vspace{1em}

\section{Monte Carlo Simulations}
We have performed Monte Carlo simulations of the model:
\begin{equation}
\mathcal{H} = - J_{\parallel} \sum_{\langle ij \rangle_\parallel} \vec{s}_i \cdot \vec{s}_j - J_{\perp} \sum_{\langle ij \rangle_\perp} \vec{s}_i \cdot \vec{s}_j
\end{equation}
on a~$L \times L \times L$ geometry with periodic boundary conditions. For the thermalization we use the rejection-free heatbath algorithm~\cite{miyatake1986the}, making use of the Von-Mises sampler from~\cite{lin2019juliastatsdistributionsjl,besancon2021distributionsjl} that is based on~\cite{best1979efficient}. The lattice is updated in a 3D checkerboard pattern, counting one full sweep as an update of all~$L^3$ spins. During measurements heatbath updates are complemented by Wolff cluster moves~\cite{swendsen1987nonuniversal,wolff1989collective} to reduce the autocorrelation time in the critical region. Furthermore, during thermalization parallel tempering swaps~\cite{swendsen1986replica} are performed after every $100$ heatbath sweeps using MPI~\cite{byrne2021mpijl}. In general, we carry out two different types of simulation runs to probe different physical aspects of the system:
\begin{itemize}
    \item \textbf{2D to 3D crossover:} Runs span a wide range of temperatures around and below~$T_{BKT}$. Small system sizes ($L \leq 30$) are sufficient, but we use narrow spacing in~$L$. For each~$(\Delta, T, L)$ we perform $400k$ thermalization sweeps and $600k$ measurement sweeps.
    \item \textbf{Criticality:} We want to be close to true critical temperature~$T_c (\Delta)$. We need large systems sizes (up to~${L = 72}$) to extrapolate to the thermodynamic limit in pairs of~$(L,2L)$. Anticipating second-order finite-size scaling, we use $L$-dependent temperature intervals:~$T_c^{estimate} (\Delta) \pm \delta \cdot L^{- 1 / \nu}$ to capture the transition with high resolution. For~${\Delta = 0.006, 0.01}$ we perform $600k$ thermalization and $1.2m$ measurement sweeps. For ${\Delta \in [0.02, 1.0]}$ we perform $400k$ thermalization and $800k$ measurement sweeps. 
\end{itemize}
Applying a logarithmic binning procedure to the raw Monte Carlo time series, we found that $400k$ heatbath sweeps are sufficient to reach the equilibrium state. For further data analysis we use bootstrap resampling with block sizes ranging from~${600-1k}$ Monte Carlo steps, having checked that the integrated autocorrelation time is significantly lower.

\section{Critical Temperature}
To find the critical temperature~$T_c$ of the model as a function of the coupling ratio~${\Delta = J_\perp / J_\parallel}$, we use the fact that, in the critical region, the Binder cumulant~$b$ and the rescaled superfluid stiffness~$j$ follow the second-order finite-size scaling:
\begin{equation}
    b(T,L)=f_b\!\left[(T-T_c)L^{1/\nu}\right], \qquad j(T,L)=f_j\!\left[(T-T_c)L^{1/\nu}\right] \label{eq:second_order_scaling}.
\end{equation}
For each quantity, we carry out the following procedure: 1)~Fix $\nu = 0.67183$ according to~\cite{xu2019highprecision}. 2)~Expand the scaling function as a Taylor polynomial of order~$k$. 3)~Perform fit on Monte Carlo data points in pairs $(L, 2L)$, keeping~$T_c$ and Taylor coefficients as free parameters. 4)~Increase the polynomial order~$k$ until the threshold~${\chi^2 / \mathrm{DOF} \leq 2}$ or maximum order of~${k = 10}$ is reached.

\begin{figure}[b!]
    \centering
    \includegraphics[width=\textwidth]{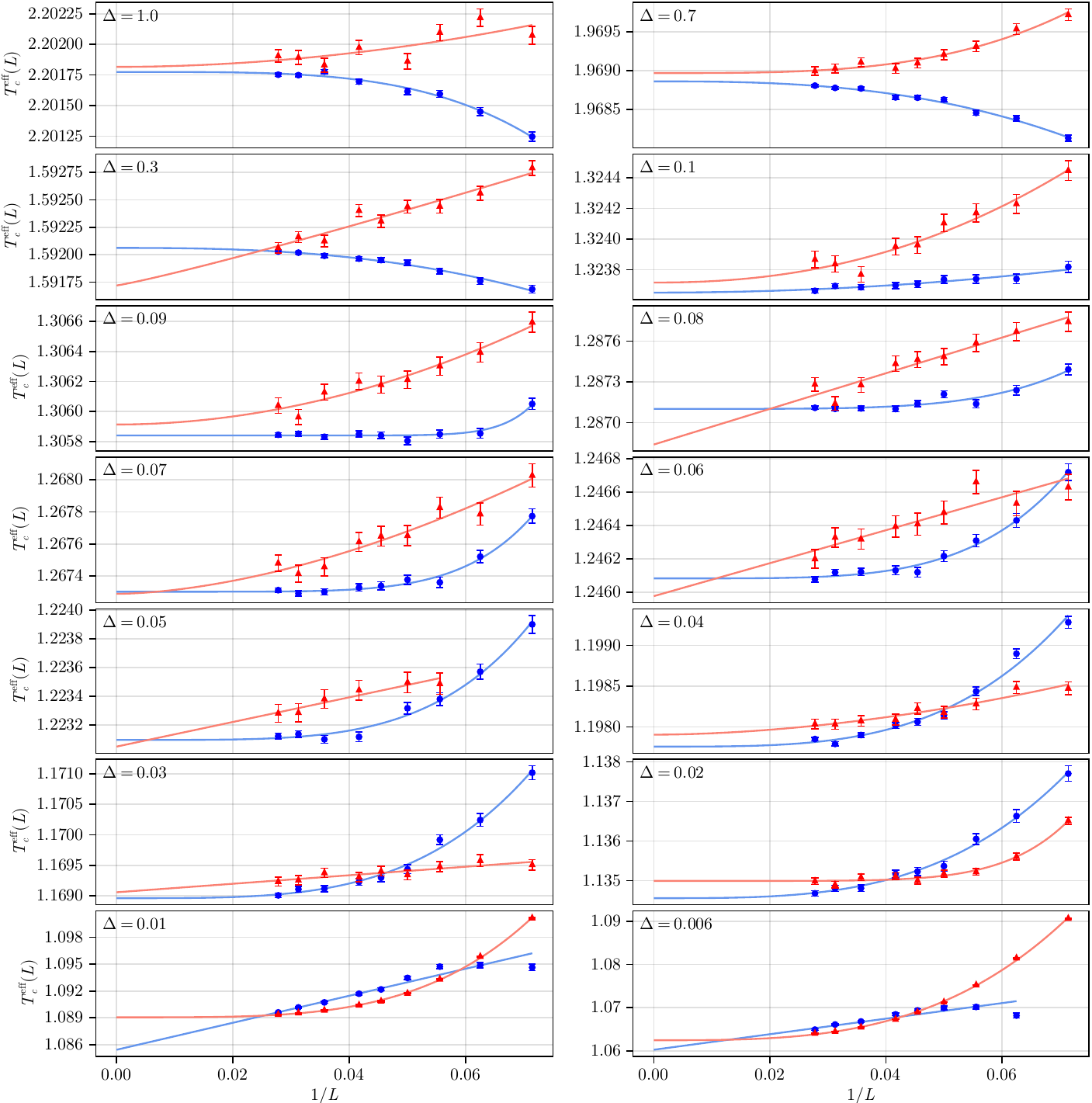}
    \caption{Effective size-dependent critical temperatures~$T_c^\mathrm{eff} (L)$ as a function of inverse system size~$1/L$: Results from fitting \cref{eq:second_order_scaling} to the Monte Carlo data are indicated by the blue dots for the Binder cumulant and red triangles for the rescaled superfluid stiffness with their associated error bars. The solid lines indicate power-law fits of the form~${T_c^\mathrm{eff} (L) = T_c + \alpha L^{-p}}$ used to extrapolate to the thermodynamic limit~${L\to\infty}$.}
    \label{fig:Tc_eff_extrapolations}
\end{figure}

By following this scheme, we obtain effective size-dependent critical temperatures~$T_c^\mathrm{eff}$ from two independent universal quantities, which we extrapolate to the thermodynamic limit with a power-law ansatz: ${T_c^\mathrm{eff} (L) = T_c + \mu L^{-p}}$, where $T_c$, $\mu$ and $p$ are treated as free fit parameters. To rule out extreme finite-size corrections from the lowest system sizes, we define the following criterion that must be satisfied before the extrapolation can be performed:
\begin{equation}
    \left[ T_c^\mathrm{eff} ( L_{\mathrm{min}} ) - T_c^{\mathrm{eff,mean}} \right] \cdot \left[ T_c^\mathrm{eff} ( L_{\mathrm{max}} ) - T_c^{\mathrm{eff,mean}} \right] \overset{!}{<} 0 ,
\end{equation}
where~$L_{\mathrm{min}}$ is the minimal system size, $L_{\mathrm{max}}$ the maximal system size and $T_c^{\mathrm{eff,mean}}$ the average of the effective critical temperature over all sizes. This criterion ensures that the data is neither too concave nor too convex to perform a good extrapolation. The smallest system sizes are systematically dropped until the criterion is satisfied. Fig. \ref{fig:Tc_eff_extrapolations} shows the effective critical temperatures together with the associated extrapolating fits for all couplings ratios~${\Delta = J_\perp / J_\parallel}$.
Finally, we combine extrapolated results from the Binder cumulant~$b$ and rescaled superfluid stiffness~$j$ by taking their average~$T_c = \sum_{\alpha=b,j} T_{c,\alpha}$ and taking into account the general systematic as well as individual statistical errors:
\begin{equation}
    {\Delta T_c}^2 = {\Delta T_{c,sys.}}^2 + {\Delta T_{c,stat.}}^2 = \frac{\sum_{\alpha = b,j} |T_{c,\alpha} - T_c|^2}{2-1} + \frac{\sum_{\alpha=b,j} \Delta {T_{c,\alpha}}^2}{2} \label{eq:Tc_error_propagation}.
\end{equation}
The final results for $T_c (\Delta)$ are tabulated in \cref{tab:critical_temperatures} with the associated error bars.
\begin{table}[t!]
    \centering
    \begin{tabular}{|l|l|}
    \hline
    $\Delta$ & $T_c(\Delta)$ \\
    \hline
    0.006 & 1.0614(29) \\
    0.01 & 1.0872(27) \\
    0.02 & 1.1348(3) \\
    0.03 & 1.1690(5) \\
    0.04 & 1.19783(16) \\
    0.05 & 1.2231(8) \\
    0.06 & 1.2460(4) \\
    0.07 & 1.26730(12) \\
    0.08 & 1.2870(4) \\
    0.09 & 1.30588(11) \\
    0.1 & 1.32368(9) \\
    0.3 & 1.5919(4) \\
    0.7 & 1.96892(9) \\
    1.0 & 2.20180(11) \\
    \hline
    \end{tabular}
    \caption{Critical temperature~$T_c$ as a function of the coupling ratio~${\Delta = J_\perp / J_\parallel}$: At each~$\Delta$ we have extrapolated~$T_c$ to the thermodynamic limit independently for both the Binder cumulant $b$ and rescaled superfluid stiffness $j$. For the final result we have computed their average and propagated errors with~\cref{eq:Tc_error_propagation}.}
    \label{tab:critical_temperatures}
\end{table}

\section{Josephson Length Scale}
\begin{figure}[b!]
    \centering
    \includegraphics[width=\textwidth]{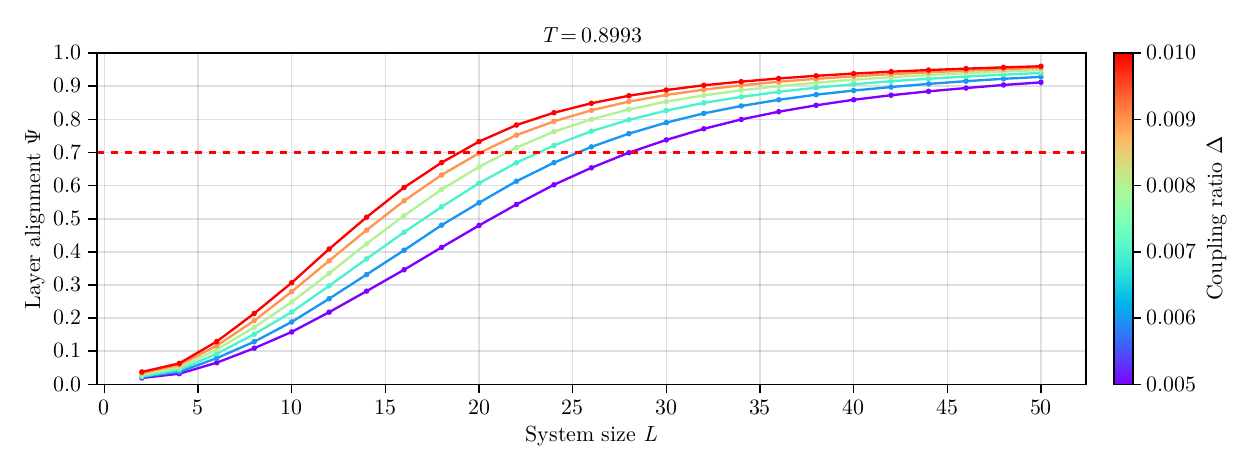}
    \caption{Layer alignment~$\Psi$ as a function of the system size~$L$: We plot how the layer alignment increases monotonically with the system size for fixed temperature~$T=0.8993 \approx T_{BKT}$ at various coupling ratios~${\Delta = J_\perp / J_\parallel}$. We define the Josephson length scale~$\ell_J$ as the value of~$L$ at which~$\Psi$ crosses the threshold~${t=0.7}$. A lower value of~$\Delta$ is associated with a higher anisotropy and results in a slower onset of the layer alignment, thus giving a larger Josephson length~$\ell_J$.}
    \label{fig:layer_alignment}
\end{figure}

In Eq.~(7) of the main text we have introduced the layer-alignment parameter~${\Psi = \langle L^{-1} \sum_l \cos ( \phi_{l+1} - \phi_l ) \rangle}$, where~$l$ identifies a single layer of the system with associated magnetized phase~$\phi_l$. It provides a natural way to probe the 2D~to~3D crossover and increases monotonically from~$0$ (randomly magnetized planes) to~$1$ (aligned planes) with increasing system size~$L$, as illustrated in \cref{fig:layer_alignment}. We can see that a larger value of the coupling ratio~${\Delta = J_\perp / J_\parallel}$ (stronger coupling among adjacent layers) leads to an earlier onset of layer alignment. By interpolating~$\Psi$ linearly between consecutive system sizes we can compute the exact value~${L = \ell_J}$ at which a given threshold~$t$ is crossed. As shown in Eq.~(9) of the main text, changing the threshold~$t$ merely sets a multiplicative constant for~$\ell_J$ but crucially does not influence the scaling with~$\Delta$.

\section{Josephson Scaling Exponent}
As argued in Eq.~(9) of the main text, we expect the layer-alignment~$\Psi$ to follow the scaling law:
\begin{equation}
    \Psi\!(L, \Delta, T) = f_{\Psi}\!\left[ L \cdot \Delta^{a(T)} \right] \label{eq:layer_alignment_scaling}.
\end{equation}
Here we describe how the scaling exponent~$a(T)$ is systematically determined using curve collapses. Let us consider the layer-alignment at a given temperature~$T$ and denote~$\Psi_{ij}$ as the Monte Carlo measurement at coupling ratio~$\Delta = \Delta_i$ and system size~$L = L_j$. We want to find the optimal exponent~$a$, such that values~$\Psi_{ij}$ obtained for different~$\Delta$ (see \cref{fig:layer_alignment}) collapse onto a single curve when plotted over the scaling argument~${L \Delta^a}$. In other words: we minimize the spread among datasets with different~$\Delta$, which is measured by the collapse error function adopted from~\cite{bhattacharjee2001measure}:
\begin{equation}
    P(a) = \sqrt{ \frac{1}{\mathcal{N}} \sum_{i} \sum_{i^\prime \neq i} \sum_{j} \theta(i, i^\prime , j) \cdot |\Psi_{i^\prime j} - \varepsilon_{i} \left( L_{j} \cdot {\Delta_{i^\prime}}^a \right)|^2 } \label{eq:josephson_collapse_error}.
\end{equation}
It is instructive to interpret \cref{eq:josephson_collapse_error} as follows:
\begin{enumerate}
    \item First, pick a reference dataset~$i$ with coupling ratio~$\Delta = \Delta_i$.
    \item Interpolate data points~$\Psi_{ij}$ of the reference set linearly between rescaled system sizes~$L_j \cdot \Delta^a$. Label the resulting interpolating function as~$\varepsilon_i$.
    \item Compare remaining datasets with~${\Delta_{i^\prime} \neq \Delta_i}$ to the reference interpolation $\varepsilon_i$ and sum the square of the residuals. The function~$\theta(i, i^\prime, j)$ ensures that only those points are counted that fall within the reference domain.
    \item Go to 1. and repeat until every~$\Delta_i$ has been chosen as reference set once.
    \item Normalize sum of residuals by the total count~$\mathcal{N} = \sum_{i, i^\prime \neq i, j} \theta(i, i^\prime, j)$.
\end{enumerate}
\begin{figure}[b!]
    \centering
    \includegraphics[width=\textwidth]{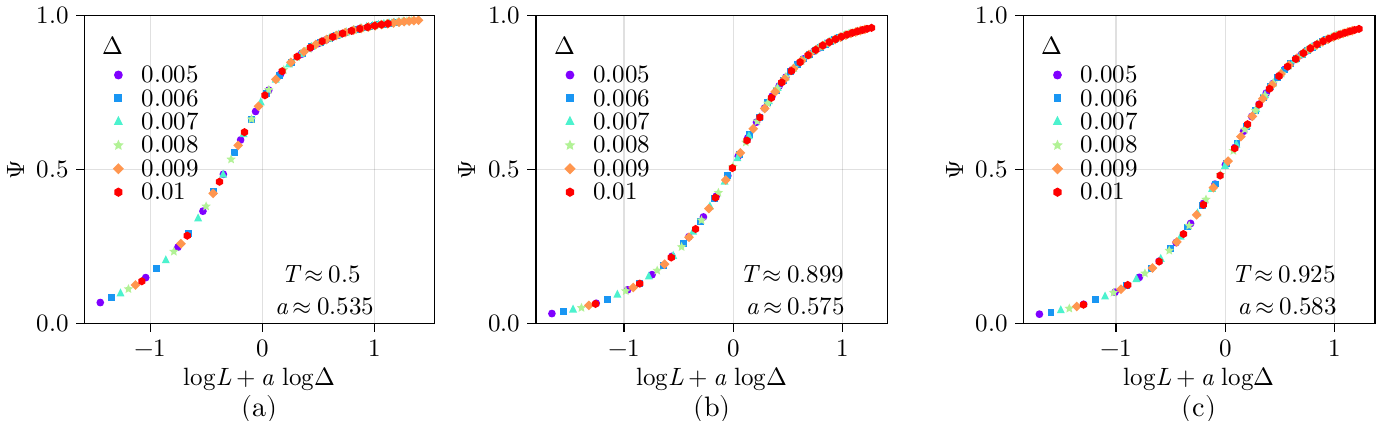}
    \caption{Curve collapse of layer-alignment~$\Psi$ for temperatures \textbf{(a)}~${T\approx0.5}$ with~${a\approx0.535}$ \textbf{(b)}~${T\approx0.899}$ with~${a\approx0.575}$ and \textbf{(c)}~${T\approx0.583}$ with~${a\approx0.583}$. We plot~$\Psi$ over the scaling argument~${\log\!\left[ L \Delta^{a (T)} \right]}$ for ${L\geq 4}$.}
    \label{fig:curve_collapses_three_temperatures}
\end{figure}
We perform a single-parameter optimization of \cref{eq:josephson_collapse_error} using~\cite{mogensen2018optim} for every bootstrap sample and each temperature to find the optimal scaling exponent~$a(T)$ with its associated statistical error. The resulting collapse is shown as an example for three different temperatures in \cref{fig:curve_collapses_three_temperatures}. While for large~$L$ the collapse works exceedingly well, there are small corrections to \cref{eq:layer_alignment_scaling} at the smallest system sizes (e.g. ${L \sim 2-10}$). This can be seen in \cref{fig:josephson_collapse_cut} for~${T = 0.8993 \approx T_{BKT}}$, where we show how the collapse error~$P$ and optimal collapse exponent~$a$ change as we exclude system sizes~${L < L_{min}}$. For each cut-off choice~${L_{min} = L_i}$ we obtain a data point~$(P_i, \eta_i , \Delta\eta_i )$, where we have used the anticipated relation~${a = (2-\eta_{2D})^{-1}}$, as introduced in Eq.~(6) of the main text, to calculate the 2D~anomalous dimension~$\eta_{2D}$ that is associated with a given collapse exponent~$a$. For the further analysis, we drop all data points with collapse error~${P_i > 0.001}$, which at~${T = 0.8993}$ for example corresponds to considering only~${L_{min} \geq 16}$ (see shaded red area in \cref{fig:josephson_collapse_cut}). Performing a full finite-size scaling analysis in this setting is difficult, so we take, as a best guess, the average of all points~$\eta_i (T)$ for each temperature as a final result. For the error bars we take into account the general systematic error as well as the individual statistical errors:
\begin{equation}
    {\Delta \eta}^2 =
    {\Delta\eta_{sys.}}^2 + {\Delta\eta_{stat.}}^2 =
    \frac{\sum_i^N | \eta_i - \eta |^2 }{N-1}
    + \frac{\sum_i^N {\Delta\eta_i}^2}{N} .
\end{equation}
\begin{figure}[t!]
    \centering
    \includegraphics[width=\textwidth]{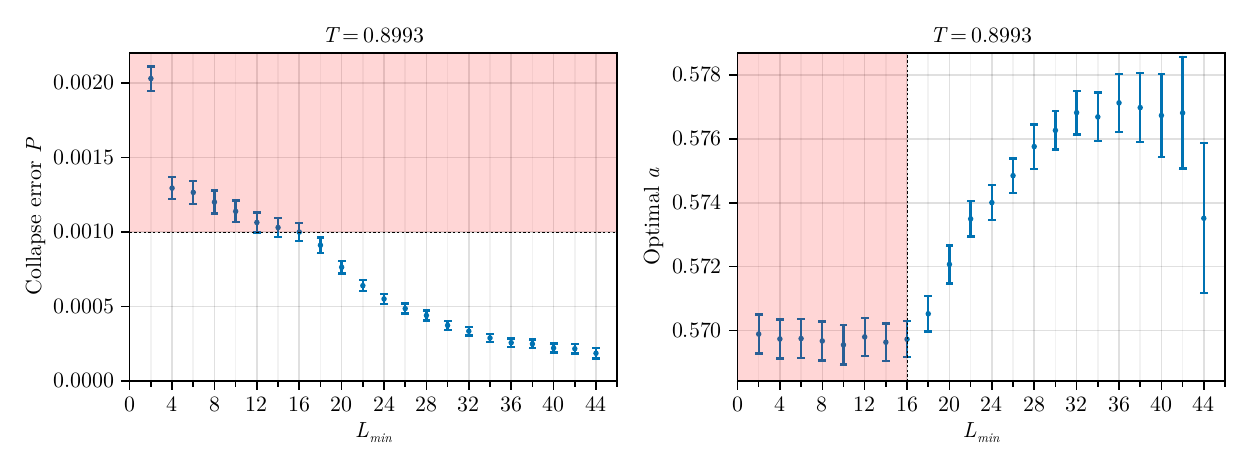}
    \caption{Curve collapse results for~$T=0.8993 \approx T_{BKT}$: We minimize the collapse error function from \cref{eq:josephson_collapse_error} for different minimum system sizes~$L_{min}$ to determine the scaling exponent in \cref{eq:layer_alignment_scaling} systematically. The left plot shows the resulting minimum collapse error~$P$ and the right plot shows the corresponding optimal collapse exponent~$a$ over~$L_{min}$. To reduce the systematic error from small system sizes, we require~$P$ to be below the threshold~$0.001$. That means in this example for~$T=0.8993$ that we disregard all data points up to~$L_{min} = 16$, as indicated by the red area in the plots.}
    \label{fig:josephson_collapse_cut}
\end{figure}

\section{BKT vs Second-Order Finite-Size Scaling}
\begin{figure}[h!]
    \centering
    \includegraphics[width=\textwidth]{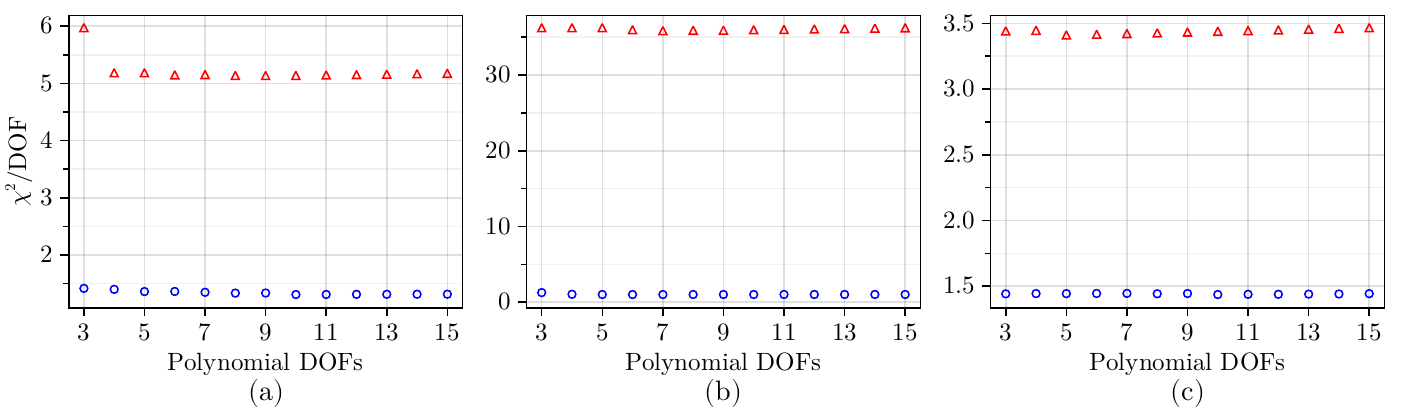}
    \caption{Goodness-of-fit~${\chi^2 / \mathrm{DOF}}$ for Binder cumulant as a function of the number of free polynomial coefficients for coupling ratios \textbf{(a)}~$\Delta = 0.006$, \textbf{(b)}~$\Delta = 0.1$ and \textbf{(c)}~$\Delta = 1$. The blue circles indicate fits assuming second-order scaling and the red triangles correspond to BKT scaling.}
    \label{fig:chi2_binder_poly_order}
\end{figure}

As described in the End Matter of the main text, we fit the Monte Carlo measurements of the Binder cumulant~$b$ and rescaled superfluid stiffness~$j$ with standard second-order as well BKT finite-size scaling and compare the resulting values of~${\chi^2 / \mathrm{DOF}}$. In Fig.~4 of the End Matter we showed that for the Binder cumulant the standard second-order scaling (see \cref{eq:second_order_scaling}) does consistently yield smaller values of~${\chi^2 / \mathrm{DOF}}$ than BKT scaling ${b(T,L) = f_b (|T-T_c| \cdot (\ln L + c)^2)}$. In \cref{tab:bkt_vs_second_order} we present the full data for both~$b$ and~$j$. Finally, we show in \cref{fig:chi2_binder_poly_order} for~$\Delta = 0.006, 0.1, 1$ as examples that~${\chi^2 / \mathrm{DOF}}$ remains stable when increasing the order of the Taylor polynomial that approximates the scaling function.
\begin{table}[b!]
\centering
\begin{tabular}{|c||c|c||c|c|}
\hline
$\Delta$ & $\chi_{b,\text{second order}}^2 / \mathrm{DOF}$ & $\chi_{b,\text{BKT}}^2 / \mathrm{DOF}$ & $\chi_{j,\text{second order}}^2 / \mathrm{DOF}$ & $\chi_{j,\text{BKT}}^2 / \mathrm{DOF}$ \\
\hline
0.006 & 1.36 & 5.13 & 446.32 & 170.71 \\
0.01 & 1.95 & 4.6 & 92.3 & 45.64 \\
0.02 & 1.44 & 2.05 & 2.55 & 3.92 \\
0.03 & 1.87 & 2.81 & 1.16 & 2.2 \\
0.04 & 1.94 & 5.92 & 1.05 & 3.55 \\
0.05 & 1.41 & 3.0 & 0.98 & 1.7 \\
0.06 & 1.34 & 9.53 & 1.05 & 3.18 \\
0.07 & 1.26 & 15.44 & 1.1 & 4.0 \\
0.08 & 1.14 & 20.97 & 1.13 & 4.31 \\
0.09 & 1.06 & 29.02 & 1.09 & 5.29 \\
0.1 & 1.02 & 35.81 & 1.11 & 6.28 \\
0.3 & 1.34 & 51.08 & 1.2 & 8.6 \\
0.7 & 1.85 & 74.07 & 1.14 & 13.65 \\
1.0 & 1.44 & 3.41 & 0.98 & 1.29 \\
\hline
\end{tabular}
\caption{Goodness-of-fit~${\chi^2 / \mathrm{DOF}}$ as a function of the coupling ratio~${\Delta = J_\perp / J_\parallel}$: We have fitted Monte Carlo data for the Binder cumulant~$b$ and rescaled superfluid stiffness~$j$ with second-order and BKT finite-size scaling forms with the critical temperature~$T_c$ and $\nu$~respectively~$c$ unconstrained. The scaling functions have been expanded as Taylor polynomials with~$6$ free coefficients. For~$b$, second-order scaling consistently outperforms BKT scaling for all~${\Delta>0}$. For $j$, the same trend holds for all~$\Delta$ except at~${\Delta=0.006}$ and~$0.01$, where both values are extraordinarily large; in this regime neither fit is acceptable and a relative comparison is not meaningful. We therefore treat these two points as outliers and base our conclusions on the remaining~$\Delta$, which favor second-order scaling.}
\label{tab:bkt_vs_second_order}
\end{table}

\section{Correlation Length Critical Exponent}
\begin{figure}[t!]
    \centering
    \includegraphics[width=0.92\textwidth]{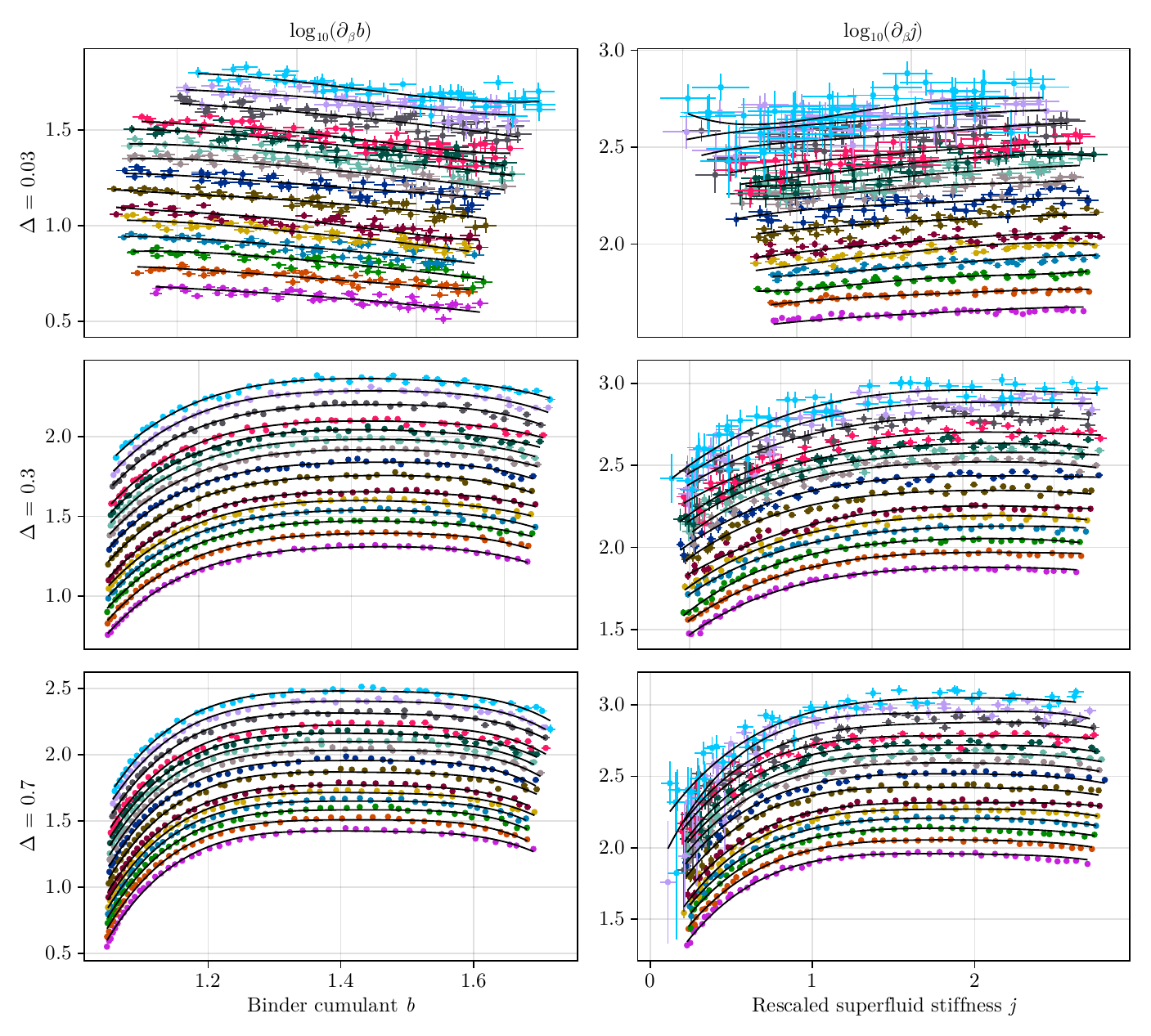}
    \caption{Derivative of Binder cumulant~$b$ and rescaled superfluid stiffness~$j$ with respect to inverse temperature~$\beta = 1/T$ for example coupling ratios~${\Delta = 0.03, 0.3, 0.7}$. Data points are indicated by circles for system sizes:~${L = 72, 64, 56, 48, 44, 40, 36, 32, 28, 24, 22, 20, 18, 16, 14}$ (from top to bottom). Each fit yields an effective size-dependent correlation length critical exponent~$\nu_{\mathrm{eff}} (L)$ and is indicated by a solid line.}
    \label{fig:nu_raw_multiplot}
\end{figure}

\begin{figure}[h!]
    \centering
    \includegraphics[width=0.92\textwidth]{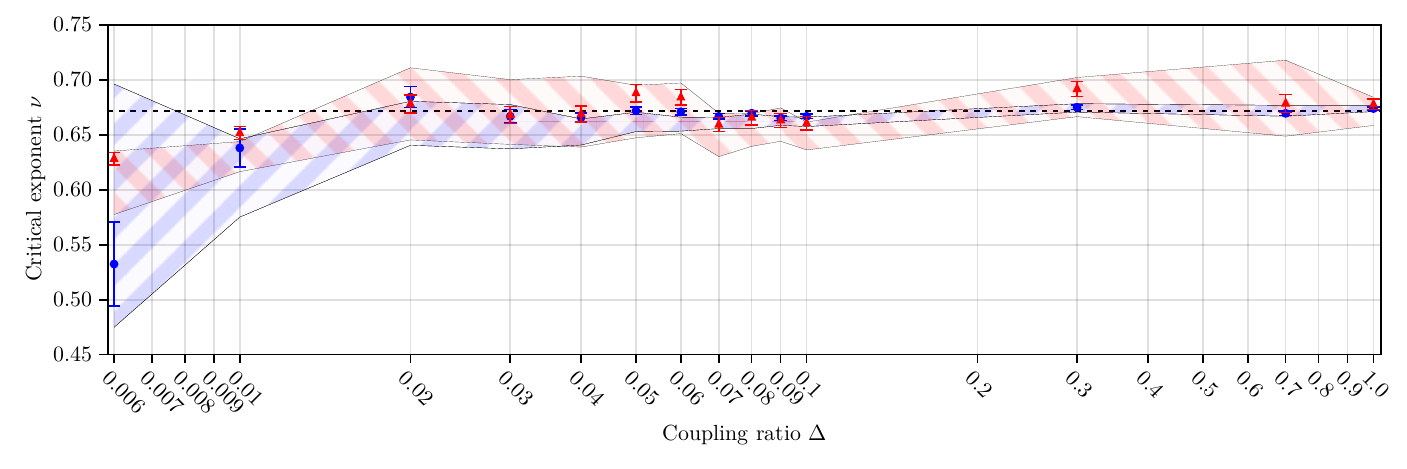}
    \caption{Correlation length critical exponent~$\nu$ as a function of the coupling ratio~${\Delta = J_\perp / J_\parallel}$: Here we show the result from Fig.~2 of main text in higher resolution. The blue dots represent extrapolation results from the Binder cumulant~$b$ and the red triangles those from the rescaled stiffness~$j$. In contrast, the average of the $5$~largest system sizes (without extrapolation) is indicated by the blue $\diagup$-shaded area for~$b$ and by the red $\diagdown$-shaded area for~$j$. The horizontal dashed line represents the isotropic high-precision result from~\cite{xu2019highprecision}.}
    \label{fig:nu_overview}
\end{figure}

\begin{figure}[t!]
    \centering
    \includegraphics[width=\textwidth]{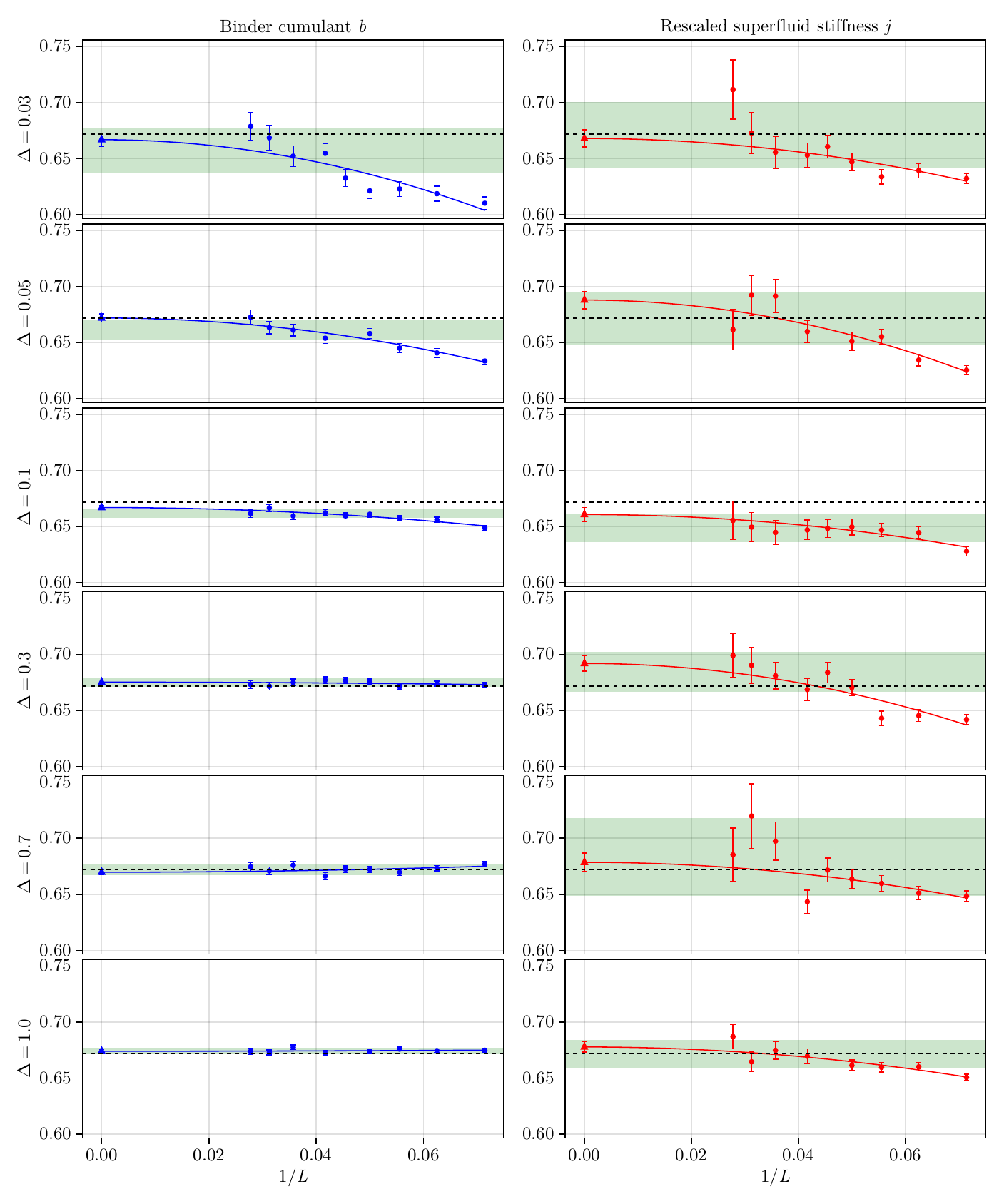}
    \caption{Extrapolation of correlation length critical exponent to thermodynamic limit for Binder cumulant~$b$ (left) and rescaled superfluid stiffness~$j$ (right): The effective size-dependent values~$\nu_{\mathrm{eff}} (L)$ (dots) are fitted with ansatz~${\nu_{\mathrm{eff}} (L) = \nu + \mu L^{-2}}$ (solid lines) to obtain the extrapolated value~$\nu$ (triangles) at~${L \to \infty}$. The green shaded area indicates the average of the~$5$ largest system sizes without extrapolation. The gray dashed line represents the high-precision literature value from~\cite{xu2019highprecision} for the isotropic model~(${\Delta = 1}$).}
    \label{fig:nu_eff_multiplot}
\end{figure}

Having established that the observed phase transition is indeed of second-order, we adopt the technique introduced in~\cite{beiming2023} to determine the correlation length critical exponent~$\nu$ and show that for any~${\Delta = J_\perp / J_\parallel}$ it flows towards the literature value of the isotropic model ($\Delta = 1$) in the thermodynamic limit. As described in Eqs.~(A.10)--(A.11) of the End Matter, we measure the derivative of the Binder cumulant~$b$ and rescaled superfluid stiffness~$j$ with respect to inverse temperature~$\beta = 1/T$:
\begin{align}
\partial_\beta b
	&=
	\frac{\partial_\beta \langle m^4 \rangle}{\langle m^2 \rangle^2}
	-
	2 \frac{\langle m^4 \rangle}{\langle m^2 \rangle^3} \partial_\beta \langle m^2 \rangle \\
	&=
	\frac{1}{\langle m^2 \rangle^2} \left( \langle m^4 \rangle \langle \mathcal{H} \rangle - \langle m^4 \mathcal{H} \rangle \right)
	-
	2 \frac{\langle m^4 \rangle}{\langle m^2 \rangle^3} \left( \langle m^2 \rangle \langle \mathcal{H} \rangle - \langle m^2 \mathcal{H} \rangle \right) \\
	&=
	- \frac{\langle m^4 \mathcal{H} \rangle}{\langle m^2 \rangle^2}
	+ 2 \frac{\langle m^2 \mathcal{H} \rangle \langle m^4 \rangle}{\langle m^2 \rangle^3}
	- \frac{\langle m^4 \rangle \langle \mathcal{H} \rangle}{\langle m^2 \rangle^2}, \\
    \partial_\beta j
	&=
	\frac{J_\parallel}{L^2} \partial_\beta \langle A \rangle - \frac{{J_\parallel}^2}{L^2} \langle B \rangle - \beta \frac{{J_\parallel}^2}{L^2} \partial_\beta \langle B \rangle \\
	&=
	\frac{J_\parallel}{L^2} \left\{ \langle A \rangle \langle \mathcal{H} \rangle - \langle A \mathcal{H} \rangle \right\}
	- \frac{{J_\parallel}^2}{L^2} \langle B \rangle - \beta \frac{{J_\parallel}^2}{L^2} \left\{ \langle B \rangle \langle \mathcal{H} \rangle - \langle B \mathcal{H} \rangle \right\} \\
	&=
	\frac{J_\parallel}{L^2} \left\{ \langle A \rangle \langle \mathcal{H} \rangle - \langle A \mathcal{H} \rangle \right\}
	- \frac{{J_\parallel}^2}{L^2} \langle B \rangle \left\{ 1 + \beta \langle \mathcal{H} \rangle \right\}
	+ \beta \frac{{J_\parallel}^2}{L^2} \langle B \mathcal{H} \rangle,
\end{align}
where we use the identity~$\partial_\beta\langle \cdot \rangle = \langle \mathcal{H} \rangle \langle \cdot \rangle - \langle \mathcal{H \cdot} \rangle$, and $A = \sum_i \vec{s}_i \cdot \vec{s}_{i+\hat{x}}$, $B = [ \sum_i {\left( \vec{s}_i \wedge \vec{s}_{i+\hat{x}} \right)_{(z)}} ]^2$ are the diamagnetic and paramagnetic contributions to the superfluid stiffness respectively. Without loss of generality we measure the superfluid stiffness along the $x$-direction of the system.
Parametrizing the derivative of each universal quantity by the quantity itself, we fit the Monte Carlo data with the ansatz~${\log \partial_\beta b = \frac{1}{\nu} \log L + \log h_b (b)}$ in pairs of~$(L, 2L)$, where we expand~$\log h_b (b)$ as a 4th~order Taylor polynomial. The results are shown in \cref{fig:nu_raw_multiplot} for both $b$~and~$j$ at example coupling ratios~${\Delta = 0.03, 0.3, 0.7}$. From this we obtain an effective size-dependent critical exponent~$\nu_{\mathrm{eff}} (L)$ that we extrapolate to the thermodynamic limit with the ansatz~${\nu_{\mathrm{eff}} (L) = \nu + \mu L^{-2}}$, keeping~$\nu$ and~$\mu$ as free fit parameters. The extrapolation to obtain the final result~$\nu (\Delta)$ (see \cref{fig:nu_overview}) is shown for example coupling ratios~${\Delta = 0.03, 0.05, 0.1, 0.3, 0.7, 1}$ in \cref{fig:nu_eff_multiplot}.
\end{document}